\documentclass[12pt]{article}
\def\cdate{March 6, 2009}
\usepackage[american]{babel}
\usepackage{latexsym,amssymb}
\usepackage{txfonts}

\makeatletter
\renewcommand{\@evenhead}{\raisebox{0pt}[\headheight][0pt]{\vbox{\hbox
to \textwidth{\thepage\hfil\strut\textit{\leftmark}}\hrule}}}
\renewcommand{\@oddhead}{\raisebox{0pt}[\headheight][0pt]{\vbox{\hbox
to \textwidth{\textit{\rightmark}\hfil\strut\thepage}\hrule}}}
\makeatother

\def\II{{\mathbb{I}}} 
\def\RR{{\mathbb{R}}} 
\def\CC{{\mathbb{C}}}

\def\tr{\mathrm{tr\,}} 
\def\Tr{\mathrm{Tr\,}} 
\def\Det{\mathrm{Det\,}} 
\def\vol{\mathrm{vol\,}}

\def\End{\mathrm{End\,}} 

\def\be{\begin{equation}}
\def\ee{\end{equation}}
\def\bea{\begin{eqnarray}}
\def\eea{\end{eqnarray}}
\def\bes{\begin{displaymath}}
\def\ees{\end{displaymath}}
\def\bmp{\begin{minipage}}
\def\emp{\end{minipage}}

\begin{document}

\begin{titlepage}
\thispagestyle{empty}
\null
\hspace*{50truemm}{\hrulefill}\par\vskip-4truemm\par
\hspace*{50truemm}{\hrulefill}\par\vskip5mm\par
\hspace*{50truemm}{{\large\sc New Mexico Tech {\rm (\cdate)}}}
\vskip4mm\par
\hspace*{50truemm}{\hrulefill}\par\vskip-4truemm\par
\hspace*{50truemm}{\hrulefill}
\par
\bigskip
\bigskip
%\par
%\hspace*{50truemm}{\LARGE\textbf{\textsf{DRAFT}}}
%\par
\vspace{2cm}
\centerline{\huge\bf Non-perturbative Effective Action}
\bigskip
\centerline{\huge\bf in Gauge Theories}
\bigskip
\centerline{\huge\bf and Quantum Gravity}
\bigskip
\bigskip
\centerline{\Large\bf Ivan G. Avramidi}
\bigskip
\centerline{\it Department of Mathematics}
\centerline{\it New Mexico Institute of Mining and Technology}
\centerline{\it Socorro, NM 87801, USA}
\centerline{\tt E-mail: iavramid@nmt.edu}
\bigskip
%\centerline{Revised on }
\medskip
%\maketitle 
\vfill

{\narrower
\par
% Abstract 

We use our recently developed algebraic methods for the calculation of the heat
kernel on homogeneous bundles over symmetric spaces to evaluate the 
non-perturbative low-energy effective action in quantum general
relativity and Yang-Mills gauge theory in curved space. We obtain an exact
integral repesentation for the effective action that generates all terms in
the standard asymptotic epxansion of the effective action without derivatives
of the curvatures effectively summing up the whole infinite subseries of all
quantum corrections with low momenta.

\par}
\vfill

\end{titlepage}

%=================================================================
%===============================================================
\section{Introduction}
\setcounter{equation}0

One of the basic object in quantum field theory is the effective action
(see, \cite{dewitt03,birrel82,avramidi91,avramidi09,avramidi00}). It is
a functional of the background fields that encodes, in principle, all the
information of quantum field theory. It determines the full one-point
propagator as well as all full vertex functions. Moreover, it gives the
effective equations for the background fields, which makes it possible to study
the back-reaction of quantum processes on the classical background. 

One of the most powerful methods for the evaluation of the effective action is
the heat kernel approach (see the books
\cite{dewitt03,birrel82,gilkey95,berline92,hurt83,avramidi00,kirsten01} and
reviews \cite{avramidi91,avramidi99a,avramidi02,avramidi99,vassilevich03}). Of
course, the effective action (or the heat kernel) cannot be computed exactly.
Therefore, various approximation schemes have been developed depending on the
problem one is studying. First of all, there is the standard semi-classical
expansion of the effective action in inverse powers of a (large) mass parameter
of massive quantum fields, which corresponds to the short-time asymptotic
expansion of the trace of the heat kernel in powers of the proper time. It
describes such physical effects as polarization of vacuum of massive quantum
fields by weak background fields. There has been tremendous progress in the
explicit calculation of the coefficients of this asymptotic expansion over the
last two decades (see,
\cite{gilkey95,kirsten01,vassilevich03,avramidi91,avramidi99,avramidi00,
avramidi02,avramidi09}). However, the applicability of this approximation is
rather limited---it does not apply to strong background fields and massless (or
light) quantum fields. Therefore, there is a need for new non-perturbative
approximation schemes.

Next, one is interested in scattering processes of
energetic particles. Such processes are well described by the (essentially
perturbative) high-energy approximation. The high-energy effective action can
be computed in a sufficiently elaborated perturbation theory
\cite{avramidi91,avramidi00,avramidi09,avramidi99a,avramidi02}. Although it is
non-local, it is analytic in the background fields and, therefore, can be
computed simply by expanding in powers of background fields (or their
curvatures). 

On another hand, one is interested in studying the structure of the physical
vacuum (the ground state) of the theory. Such problems are well described by
the low-energy approximation. The low energy effective action (or the heat
kernel) is a local, but highly non-trivial (non-polynomial) functional of
background fields and their curvatures, and, therefore, it cannot be computed
in the usual perturbation theory. There are just a few very special cases, such
as group manifolds, spheres, rank-one symmetric spaces and split-rank symmetric
spaces when one can determine the spectrum of the Laplacian exactly and obtain
closed formulas for the heat kernel in terms of the root vectors and their
multiplicities \cite{camporesi90,hurt83}. The complexity of the method
crucially depends on the global structure of the symmetric space, most
importantly its rank. Therefore, to study the low-energy effective action in
the generic case one needs new essentially non-perturbative methods. The
development of such methods for the calculation of the heat kernel was
initiated in our papers \cite{avramidi93,avramidi95c} for a gauge theory in
flat space, which were then applied to study the vacuum structure of the
Yang-Mills theory in \cite{avramidi95b,avramidi99}. These ideas were first
extended to scalar fields on curved manifolds in \cite{avramidi94a,avramidi96}
and finally to arbitrary twisted spin-tensor fields in
\cite{avramidi08a,avramidi08b}.

In the present paper we apply these methods to study the one-loop low-energy
effective action in quantum general relativity and Yang-Mills theory in curved
space with some twisted scalar and spinor fields.
We consider a wide class of field theory models with the action
\bea
S&=&\int\limits_M dx\,g^{1/2}\Biggl\{
\frac{1}{k^2}(R-2\Lambda)
+
{1\over 8e^2}{\rm tr}\,{\cal F}_{\mu\nu}{\cal F}^{\mu\nu}
\nonumber\\[5pt]
&&
+\bar\psi\left[\gamma^\mu\nabla_\mu
+M_{}(\varphi)\right]\psi  
-{1\over 2}g^{\mu\nu}\nabla_\mu\bar\varphi\nabla_\nu\varphi
-V(\varphi)\Biggr\},
\label{(1)}
\eea
where $g=|\det g_{\mu\nu}|$, $g_{\mu\nu}$ is a metric on the spacetime manifold
$M$, $R$ is the scalar curvature, $k^2=16\pi G$ is the Einstein coupling
constant, $G$ is the Newtonian gravitational constant, $\Lambda$ is the
cosmological constant, $ {\cal  F}_{\mu\nu}$ is the strength of the gauge
fields ${\cal A}_\mu$ taking values in the adjoint representation of the Lie
algebra of a compact simple gauge group $G_{YM}$, $e$ is a coupling constant,
$\varphi$ and $\psi$ are multiplets of real scalar fields and the Dirac spinor
ones, which belong to some, in general, different representations of the gauge
group, $M_{}(\varphi)$ is a spinor mass matrix, $V(\varphi)$ is a potential
for scalar fields, $\gamma^\mu$ are the Dirac matrices and $\nabla_\mu$ is the
covariant derivative in the corresponding representation.

Our goal is to compute the one-loop effective action for this model assuming
a covariantly constant background, that is, a background  metric with
covariantly constant curvature, a background gauge field with the covariantly
constant strength tensor and also some covariantly constant background scalar
fields. 

This paper is organized as follows. In Sec. 2 we describe briefly the
construction of the one-loop effective action in gauge field theories.
In Sec. 3. we describe the heat kernel method for the calculation of functional
determinants of elliptic partial differential operators of Laplace type.
In Sec. 4. we describe the low-energy approximation and derive some of
its consequences, in particular, we present the results for the heat trace of
our earlier paper \cite{avramidi08b}. In Sec. 5-7 we apply these results
to evaluate the effective action in general relativity, the Yang-Mills
theory and also for the matter (scalar and the spinor) fields. In conclusion
we summarize our results.

%=========================================================
\section{Effective Action in Gauge Field Theories}

We describe briefly the construction of the one-loop effective action in gauge
field theories.
Let $M$ be a globally hyperbolic spacetime manifold with a 
(pseudo)-Rie\-man\-nian metric.
Let ${\cal V}$ and ${\cal G}$ be two fiber bundles over $M$ such that
$\dim{\cal G}<\dim{\cal V}$.
Let both bundles $V$ and $G$ be equipped with some Hermitian 
positive-definite metrics
and with the corresponding natural $L^2$ scalar products
$(\;,\;)_{\cal V}$ and $(\;,\;)_{\cal G}$.

The sections of the bundle ${\cal V}$ are quantum (gauge) fields.
The dynamics of the quantum fields is described by the action
$S: C^\infty({\cal V})\to \RR$.
At the linearized level it is described by the
second order differential operator,
$
P:\ C^\infty(T{\cal V})\to C^\infty(T{\cal V})
$
defined by
\be
(h,Ph)_V=\frac{d^2}{d\varepsilon^2}
S(\varphi+\varepsilon h)\Big|_{\varepsilon=0}\,.
\ee
If this operator is non-degenerate then the one-loop effective action is
determined by its determinant \cite{dewitt03}
\bea
\Gamma_{(1)} &=&\sigma\frac{i}{2}\log\Det\,(-P)\,.
%\nonumber
\eea
where $\sigma=+1$ for bosonic fields and $\sigma=-1$ for fermionic fields.
In the following we consider the bosonic theory.

In gauge theory the operator $P$ is degenerate. This means that the action
has
some invariant flows which define a first order differential
operator
$N: C^\infty(T{\cal G})\to C^\infty(T{\cal V})$.
Let $\bar N: C^\infty(T{\cal V})\to C^\infty(T{\cal G})$ be the first order
differential operator such that for any $\xi\in C^\infty({\cal G})$, $h\in
C^\infty(T{\cal V})$ 
\be
(\bar Nh,\xi)_{\cal G} = (h,N\xi)_{\cal V}\,,
\ee
and $F: C^\infty(T{\cal G})\to C^\infty(T{\cal G})$ be the operator defined by
\be
F=\bar NN.
% \qquad H=R\bar R.
\ee
Finally, let $L: C^\infty(T{\cal V})\to C^\infty(T{\cal V})$ be the
second-order differential operator defined by
\be
L=-P-N\bar N
\,.
\ee

We consider only the case when the gauge generators are linearly independent.
This means that the rank of the leading symbol of the operator $N$ equals the
dimension of the bundle ${\cal G}$. 
We also assume that the leading symbols of
the
generators $N$ are complete in that they generate all zero-modes of the leading
symbol of the operator $P$. Then the leading symbols of the operators $L$ and
$F$ are non-degenerate and  the one-loop effective action has the form
\cite{dewitt03,avramidi00}
\bea
\Gamma_{(1)} &=& {i\over 2 }
\Big(\log\Det\,L-2\log\Det\,F\Big)\,.
%\nonumber
\eea
Strictly speaking, one should include the contribution, $\log\Det\gamma$, of
the determinant of the gauge group metric $\gamma$. 
However, in the cases of our primary interest
(general relativity and Yang-Mills theory) the gauge group metric $\gamma$ is a
zero-order differential operator, and, therefore, its contribution can be
omitted, more precisely, it can be absorbed in the definition of the path
integral measure.

%==========================================================
%%%%%%%%%%%%%%%%%%%%%%%%%%%%%%%%%%%%%%%%%%%%%%%%%%%%%%%%%%%

\section{Heat Kernel Method}
\setcounter{equation}0

The effective action is determined by the functional determinants of
second-order hyperbolic partial differential operators with Feynman boundary
conditions. At this point we can do the analytic continuation to the imaginary
time (Wick rotation) and consider instead of hyperbolic operators the elliptic
ones. Furthermore, the most important elliptic partial differential operators
encountered in quantum field theory are so-called Laplace type operators. That
is why we concentrate below on the calculation of the heat kernel for Laplace
type operators (see
\cite{gilkey95,berline92,avramidi00,avramidi02,avramidi09}).

%=====================================================
%\subsection{Laplace Type Operators}

Let $(M,g)$  be a smooth compact Riemannian manifold of dimension $n$ without
boundary, equipped with a positive definite Riemannian metric $g$. We assume
that it is complete simply connected orientable and spin. Let $\Lambda$ be a
vector space and $\End(\Lambda)$ be the space of endomorphisms of $\Lambda$.
Let $\mathcal{T}$ be a spin-tensor bundle with fiber $\Lambda$ realizing a
representation of the spin group ${\rm Spin}(n)$. It naturally defines a
representation $\Sigma: {\cal SO}(n)\to \End(\Lambda)$ of the orthogonal
algebra ${\cal SO}(n)$ in $\Lambda$ with generators $\Sigma_{ab}$. The spin
connection induces a connection on the bundle $\mathcal{T}$ defining the
covariant derivative of spin-tensor fields.

Let $G_{YM}$ be a  compact Lie (gauge) group and ${\cal G}_{YM}$ be its Lie
algebra. It naturally defines the principal fiber bundle over the manifold $M$ 
with the structure group $G_{YM}$. Let $W$ be a vector space and $\End(W)$ be
the space of its endomorphisms. We consider a representation $X: {\cal
G}_{YM}\to \End(W)$ of the Lie algebra ${\cal G}_{YM}$ in $W$ and the
associated vector bundle ${\cal W}$ through this representation with the same
structure group $G_{YM}$ whose typical fiber is $W$. Then for any spin-tensor
bundle $\mathcal{T}$ we define the twisted spin-tensor bundle $\mathcal{V}$ via
the twisted product of the bundles $\mathcal{W}$ and $\mathcal{T}$ with the
fiber $V=\Lambda\otimes W$.

We assume that the vector bundle ${\cal V}$ is equipped with a Hermitian
metric. This naturally identifies the dual vector bundle ${\cal V}^*$ with
${\cal V}$. We assume that the connection $\nabla$ is compatible with the
Hermitian metric on the vector bundle ${\cal V}$. The connection is given its
unique natural extension to bundles in the tensor algebra over ${\cal V}$ and
${\cal V}^*$.  In fact, using the spin connection together with the
connection on the bundle ${\cal V}$, we naturally obtain connections on all
bundles in the tensor algebra over ${\cal V},\,{\cal V}^*,\,TM$ and $T^*M$; the
resulting connection will usually be denoted just by $\nabla$. It is usually
clear which bundle's connection is being referred to, from the nature of the
section being acted upon.

%========================================

Let $\mathcal{A}$ be a connection one form on the bundle $\mathcal{W}$ (called
Yang-Mills or gauge connection) taking values in the Lie algebra
$\mathcal{G}_{YM}$. Then for any section of the bundle ${\cal V}$ we have
\be
[\nabla_\mu,\nabla_\nu]\varphi
={\cal R}_{\mu\nu}\varphi\,,
\label{219mm}
\ee
where
\be
{\cal R}_{\mu\nu}=
\frac{1}{2}R^{ab}{}_{\mu\nu} \Sigma_{ab}
+X({\cal F}_{\mu\nu})\,,
\label{220}
\ee
is the curvature of the total connection on the
bundle $\mathcal{V}$, and 
\be
\mathcal{F}_{\mu\nu}
=\partial_\mu\mathcal{A}_\nu
-\partial_\nu\mathcal{A}_\mu
+[\mathcal{A}_\mu,\mathcal{A}_\mu]
\ee
is the curvature of the Yang-Mills connection.
We use Greek indices to denote tensor components in the coordinate basis. We
also use Latin indices from the beginning of the alphabet to denote the indices
of an orthornomal frame. Both group of indices 
range over $1,\dots n$.

%==============================================

The fiber inner product on the bundle ${\cal V}$ defines a natural $L^2$ inner
product on $C^\infty({\cal V})$. The completion of $C^\infty({\cal V})$ in this
norm defines the Hilbert space $L^2({\cal V})$. Let $\nabla^*$ be the formal
adjoint to $\nabla$ and $Q$ be a smooth endomorphism of the bundle ${\cal V}$.
A Laplace type operator $L: C^\infty({\cal V})\to C^\infty({\cal V})$ is a
partial differential operator of the form  
\be 
L=\nabla^*\nabla+Q=-\Delta+Q\,.
\label{1ms} 
\ee
where $\Delta=g^{\mu\nu}\nabla_\mu\nabla_\nu$ is the covariant Laplacian.
It is easy to show that the Laplacian, $\Delta$, and,
therefore, the operator $L$, is a self-adjoint elliptic partial differential
operator \cite{gilkey95}.

%===============================================
%\subsection{Determinants of Elliptic Operators}

{}For $t>0$ the operators 
$
U(t)=\exp(-tL)
$
form a semi-group of bounded
operators on $L^2({\cal V})$, the heat semi-group. 
Moreover, the heat semigroup $U(t)$ is a trace-class operator  with a well
defined $L^2$-trace, the heat trace \cite{gilkey95}:
\be
\Theta(t)=\Tr_{L^2}\exp(-tL)\,.
\ee
The heat trace 
is well defined for real positive $t$. In fact, it can
be
analytically continued to an analytic function of $t$
in the right half-plane (for ${\rm Re}\, t>0$).

The heat trace determines the zeta-function,
\be
\zeta(s,\lambda)=\mu^{2s}\Tr_{L^2}(L-\lambda)^{-s}
={\mu^{2s}\over\Gamma(s)}\int\limits_0^\infty dt\; t^{s-1}\,
e^{t\lambda}\Theta(t),
\ee
where $\mu$ is a renormalization parameter introduced to preserve dimensions,
$\lambda$ is a sufficiently large negative constant such that the operator
$(L-\lambda)$ is positive and $s$ is a complex parameter with ${\rm Re}\,
s>n/2$. The zeta-function is a meromorphic function of $s$ analytic at
$s=0$ \cite{gilkey95}, and, therefore, it enables one to define, in
particular, the zeta-regularized determinant of the operator $(L-\lambda)$,
via \cite{avramidi91,avramidi00,avramidi02,avramidi09}
\be
\zeta'(0,\lambda)\equiv {\partial\over\partial s}\zeta(s,\lambda)\Big|_{s=0}
=-\log\Det (L-\lambda),
\ee
which determines the one-loop effective action in quantum field theory.
The parameter $\lambda$ serves here as an infrared regularization parameter.
One should take the limit $\lambda\to 0$ at the end of the calculation. 

%====================================================================
\section{Low Energy Approximation}
\setcounter{equation}0

Of course, it is impossible to compute the heat kernel in the generic case.
That is why, one considers various approximations. To study the structure of
the ground state in quantum field theory one needs to evaluate the heat kernel
in the low-energy approximation. In this case the curvatures are strong but
slowly varying, i.e. the powers of the
curvatures are more important than the
derivatives of them.
The main terms in this approximation are the terms without any covariant
derivatives
of the curvatures.
We will consider the zeroth order of this approximation
which corresponds simply to covariantly constant background
\be
\nabla_\mu R_{\alpha\beta\gamma\delta}=0\,,
\qquad
%\ee
%\be
\nabla_\mu\mathcal{R}_{\alpha\beta}=0\,,
\qquad
%\ee
%\be
\nabla_\mu Q=0\,.
\label{31xxz}
\ee
Riemannian manifolds with parallel curvature are called symmetric spaces.
Vector bundles with parallel curvature are called homogeneous bundles.
Thus, the most general covariantly constant background is described by
homogeneous vector bundles over symmetric spaces.

%================================================
%=======================
\subsection{Holonomy Group}

A generic symmetric space has the structure
$
M=M_0\times M_s\,,
$
where $M_0=\RR^{n_0}$, $M_s=M_+\times M_-$,
and $M_+$ and $M_-$ are compact and noncompact
symmetric spaces respectively \cite{wolf72,helgason84,ruse61}.
The components of the curvature tensor
can be presented in the form \cite{ruse61,avramidi94a,avramidi96}
\be
R_{abcd} = \beta_{ik}E^i{}_{ab}E^k{}_{cd}\,,
\label{236}
\ee
where $E^i{}_{ab}$ is a collection of $p$ anti-symmetric matrices and
$\beta_{ik}$ is a symmetric nondegenerate  
$p\times p$ matrix. The number $p$ is determined by the curvature tensor.
In the following the
Latin indices from the middle of the alphabet range over $1,\dots, p$ and
will be raised and lowered with
the matrix $\beta_{ik}$ and its inverse
$\beta^{ik}$.
They should not be confused with the Latin indices from
the beginning of the alphabet.

Next, we define the traceless $n\times n$
matrices $D_i=(D^a{}_{ib})$, by
\be
D^a{}_{ib}=-\beta_{ik}E^k{}_{cb}\delta^{ca}\,.
\ee
The matrices $D_i$ are known to be the generators of the  holonomy
algebra, $\mathcal{H}$, i.e. the Lie algebra of the restricted holonomy 
group, $H$, \cite{wolf72,ruse61,avramidi96}
\be
[D_i,D_j]=F^k{}_{ij}D_k\,,
\label{43ccz}
\ee
where $F^j{}_{ik}$ are the structure constants.

The holonomy algebra is a subalgebra of the orthogonal algebra ${\cal SO}(n)$
\cite{wolf72,barut77,takeuchi91}.
The embedding of the holonomy algebra ${\cal H}$ in the orthogonal algebra
${\cal SO}(n)$ is described as follows \cite{avramidi08b}. 
Let $Y_{ab}$ be the generators of the orthogonal algebra $\mathcal{SO}(n)$ in
the representation $Y: {\cal SO}(n)\to \End(W)$ of the 
orthogonal algebra $\mathcal{SO}(n)$ in a vector space $W$ 
and let $T_i$ be the
matrices defined by
\be
T_i=-\frac{1}{2}D^a{}_{ib}Y^b{}_a\,.
\label{45ccz}
\ee
Then $T_i$ form a representation of the holonomy algebra ${\cal H}$ in $W$,
that is, they satisfy the commutation relations 
\be
[T_i,T_j]=F^k{}_{ij}T_k\,.
\label{43ccx}
\ee
Vice versa, for every representation 
$T: {\cal H}\to \End(W)$
of the holonomy algebra ${\cal H}$ in a vector space $W$ there is a
representation $Y: {\cal SO}(n)\to \End(W)$  of the 
orthogonal algebra $\mathcal{SO}(n)$ in $W$ such that the generators $T_i$ of 
the representation $T$ are given by (\ref{45ccz}).

The structure constants $F^j{}_{ik}$
of the holonomy group define the
$p\times p$ matrices $F_i$, by 
$
(F_i)^j{}_k=F^j{}_{ik}\,,
$
which generate
the adjoint representation  of the holonomy algebra.
The scalar curvature of the holonomy group is given by the invariant
\cite{wolf72,helgason84}
\be
R_H=-\frac{1}{4}\beta^{ij}F^k{}_{il}F^l{}_{jk}\,.
\ee

%=========================================================
%===============================================================
%==============================================================
\subsection{Homogeneous Vector Bundles}

Let $h^a{}_b$ be the projection 
to
the subspace $T_x M_s$ of the tangent space $T_xM$
and
\be
q^a{}_{b}=\delta^a{}_b-h^a{}_b\,
\ee
be the projection tensor to the flat subspace $\RR^{n_0}$.
Since the curvature exists only in the semi-simple submanifold $M_s$, the
components of the curvature tensor $R_{abcd}$, as well as the tensors
$E^i{}_{ab}$,  are non-zero only in the semi-simple subspace $M_s$.
Moreover, the condition (\ref{31xxz}) 
imposes strong constraints on the curvature of the homogeneous bundle
${\cal W}$.
We decompose the Yang-Mills curvature according to
\be
\mathcal{F}_{ab}
=\mathcal{B}_{ab}+\mathcal{E}_{ab}\,,
\ee
where
\be
\mathcal{B}_{ab}=\mathcal{F}_{cd}q^c{}_a q^d{}_b\,,
\qquad
\mathcal{E}_{ab}=\mathcal{F}_{cd}h^c{}_a h^d{}_b\,.
\label{2139}
\ee

Then, one can show \cite{avramidi08b} that if  $\mathcal{B}_{ab}$ is non-zero
then it takes values in an Abelian ideal  of the gauge algebra
$\mathcal{G}_{YM}$ (that is, commutes with everyhting else) and if
$\mathcal{E}_{ab}$ is non-zero then it takes values in a representation of the
holonomy algebra. More precisely, the existence of a non-zero component ${\cal
E}_{ab}$ is possible only if the holonomy algebra $\mathcal{H}$ is an ideal of
the gauge algebra $\mathcal{G}_{YM}$. That is, the gauge algebra ${\cal
G}_{YM}$ must be big enough to have a subalgebra ${\cal C}\oplus{\cal H}$,
where ${\cal C}$ is an Abelian ideal. Below we will assume that this is
the
case. 

Since the curvature ${\cal E}_{ab}$ takes values in the holonomy algebra, it
has the form \cite{avramidi08b} $X({\cal E}_{ab})=-E^i_{ab}T_i$, where $T_i$
are the generators of
the holonomy algebra in some representation $T$ of the holonomy algebra in the
vector space $W$. Since the holonomy algebra is a subalgebra of the orthogonal
algebra ${\cal SO}(n)$ it can be embedded in the orthogonal algebra ${\cal
SO}(n)$ via a representation $Y: {\cal SO}(n)\to \End(W)$
of the orthogonal algebra ${\cal SO}(n)$ in $W$,  eq.
(\ref{45ccz}). Thus, the curvature of the homogeneous bundle $\mathcal{W}$ is
given by
\bea
X(\mathcal{F}_{ab})&=&
-E^i{}_{ab}T_i+X(\mathcal{B}_{ab})
=\frac{1}{2}R^{cd}{}_{ab}Y_{cd}
+X(\mathcal{B}_{ab})
\,,
\label{2153mmz}
\eea
where $X({\cal B}_{ab})$ satisfies the commutation relations 
$[X({\cal B}_{ab}),X({\cal B}_{cd})]=[X({\cal B}_{ab}),Y_{cd}]=0$ and 
$Y_{ab}$ are the generators of the orthogonal algebra ${\cal SO}(n)$ in the
representation $Y$.

Now, we 
consider the representation $\Sigma: {\cal SO}(n)\to \End(\Lambda)$ of
the orthogonal algebra
${\cal SO}(n)$ in the vector space $\Lambda$
(defining the spin-tensor bundle $\mathcal{T}$) and
define the 
generators 
\be
G_{ab}=\Sigma_{ab}\otimes\II_Y+\II_\Sigma\otimes Y_{ab}\,
\ee
of the orthogonal algebra
${\cal SO}(n)$
in the product representation $G=\Sigma\otimes Y: 
{\cal SO}(n)\to \End(V)$ in the vector space $V=\Lambda\otimes W$.

Then the matrices
\be
{\cal R}_i=-\frac{1}{2}D^a{}_{ib}G^b{}_a
\ee
form a representation ${\cal R}: {\cal H}\to \End(V)$ of the holonomy algebra
in $V$
and the total curvature of  the
twisted spin-tensor bundle $\mathcal{V}$ is
\bea
{\cal R}_{ab}&=&
-E^i{}_{ab}{\cal R}_i+X({\cal B}_{ab})
=
\frac{1}{2}R^{cd}{}_{ab}G_{cd}+X(\mathcal{B}_{ab})
\,.
\label{2144}
\eea
The Casimir operator of the holonomy group in this representation is
\cite{avramidi08b}
\be
{\cal R}^2=\beta^{ij}{\cal R}_i{\cal R}_j
=\frac{1}{4}R^{abcd}G_{ab}G_{cd}\,.
\ee

%============================================================
\subsection{Heat Trace}
%=============================================================
%======================================================
%======================================================

The heat trace of the operator $L$ was computed in \cite{avramidi08b}.
It has the form
\bea
\Theta(t)
&=&
\int\limits_M dx\,g^{1/2}\;
(4\pi t)^{-n/2}
\exp\left\{\left({1\over 8} R + {1\over 6} R_H\right)t\right\}
\nonumber\\[10pt]
&&
\times
\int\limits_{\RR^n_{\rm reg}} 
\frac{d\omega}{(4\pi t)^{p/2}}\;\beta^{1/2}
\exp\left\{-{1\over 4 t}\left<\omega,\beta\omega\right>\right\}
\Psi(t,\omega)
\nonumber\\[10pt]
&& 
\times\left[\det{}_\mathcal{H}
\left({\sinh\left[\,F(\omega)/2\right]\over 
\,F(\omega)/2}\right)\right]^{1/2}
\left[\det{}_{TM}\left({\sinh\left[\,D(\omega)/2\right]\over 
\,D(\omega)/2}\right)\right]^{-1/2}
\label{437a}
\eea
where $\beta=\det \beta_{ij}$, 
$\left<\omega,\beta\omega\right>=\beta_{ij}\omega^i\omega^j$,
and
\be
\Psi(t,\omega)=
\tr_W
\left[\det{}_{TM}\left(\frac{
\sinh(tX(\mathcal{B}))}{tX(\mathcal{B})}\right)\right]^{-1/2}
\tr_\Lambda\exp\left[-t\left({\cal R}^2+Q\right)\right]
\exp\left[\mathcal{R}(\omega)\right]\,.
\label{314xxz}
\ee
Here $D(\omega)$, $F(\omega)$, ${\cal R}(\omega)$ and ${\cal B}$ are matrices
defined by $D(\omega)=\omega^i D_i$, $F(\omega)=\omega^i F_i$, ${\cal
R}(\omega)=\omega^i {\cal R}_i$ and ${\cal B}=({\cal B}^a{}_{b})$, where the
matrices $D_i$, $F_i$, ${\cal R}_i$ and ${\cal B}_{ab}$ were defined above in
sec. 3.1 and 3.2. Notice that the whole structure of this expression is the
same for all vector bundles (all representations), the only difference is in
the function $\Psi(t,\omega)$.

%===========================================================
%\paragraph{Remarks.}

We need to explain the meanning of the integral  over $\omega^i$ in
(\ref{437a}). In the derivation of this formula in \cite{avramidi08b} we used a
certain regularization procedure. The point is that the integrals over the
holonomy group in canonical coordinates $\omega^i$ have singularities that need
to be avoided (or regularized) by deforming the contour of integration. This
procedure with the nonstandard contour of integration is necessary  for the
convergence of the integrals since we are treating both the compact and the
non-compact symmetric spaces simultaneously. We complexify the holonomy group
by extending the canonical coordinates $\omega^i$ to be complex, more
precisely, to take values in the $p$-dimensional subspace $\RR^p_{\rm reg}$ of
$\CC^p$ obtained by rotating $\RR^p$ counterclockwise by $\pi/4$ in $\CC^p$,
that is, we replace each variable $\omega^j$ by $e^{i\pi/4}\omega^j$. We also
make an analytic continuation in the complex plane of $t$ with a cut along the
negative imaginary axis so that $-\pi/2<\arg\,t<3\pi/2$ and consider $t$ to be
real negative, $t<0$. Remember, that, in general, the nondegenerate diagonal
matrix $\beta_{ij}$ is not positive definite. The space $\RR^p_{\rm reg}$ is
chosen in such a way to make the Gaussian exponent purely imaginary. Then the
indefiniteness of the matrix $\beta$ does not cause any problems. Moreover, the
integrand does not have any singularities on these contours. The convergence of
the integral is guaranteed by the exponential growth of the sine for imaginary
argument. These integrals can be computed then in the following way. The
coordinates $\omega^j$ corresponding to the compact directions  are rotated
further by another $\pi/4$ to imaginary axis and the coordinates $\omega^j$
corresponding to the non-compact directions are rotated back to the real axis.
Then, for $t<0$ all the integrals are well defined and convergent and define an
analytic function of $t$ in a complex plane with a cut along the negative
imaginary axis.

%=======================================================
%=======================================================
\section{General Relativity}
\setcounter{equation}0

Einstein's theory of general relativity is a gauge theory with
the gauge group ${\cal G}$ being the group of diffeomorphisms of the
spacetime manifold $M$. 
The  gravitational field can be parametrized by the 
metric tensor of the space-time
$g_{\mu\nu}$.
The Hilbert-Einstein action of general relativity
has the form  
\be
S_{GR}=\frac{1}{k^2}\int\limits_M dx\;g^{1/2}
\left(R-2\Lambda\right)\;.
\ee

The tangent bundle to the bundle of Riemannian metrics is the bundle
${\cal T}_{(2)}=T^*M \vee T^*M$ of symmetric covariant
2-tensors. (Here $\vee={\rm Sym}\,\otimes$ is the symmetric tensor product).
An invariant fiber metric on the vector bundle 
${\cal T}_{(2)}$ 
is defined by
\be
E^{\mu\nu\alpha\beta}=g^{\mu(\alpha}g^{\beta)\nu}
-\varkappa g^{\mu\nu}g^{\alpha\beta}\,,
\ee
where
$\varkappa\ne 1/n$ is a real parameter.
The inverse metric is then 
\be
E^{-1}_{\mu\nu\alpha\beta}=g_{\mu(\alpha}g_{\beta)\nu}
-\frac{\varkappa}{n\varkappa-1}g_{\mu\nu}g_{\alpha\beta}\,.
\ee
The tangent bundle to the group of diffeomorphisms is the tangent bundle
$TM$ itself. We define an invariant metric on the gauge algebra 
by
\be
\gamma_{\mu\nu}
=\frac{k^2}{\alpha}g_{\mu\nu}\;,
\ee
where $\alpha\ne 0$ is a real parameter. 

The invariant flows of the action are the infinitesimal diffeomorphisms, which
define the first order differential operators
$N: C^{\infty}(TM)\to C^\infty({\cal T}_{(2)})$ 
and $\bar N: C^\infty({\cal T}_{(2)})\to C^{\infty}(TM)$ 
by
\bea
(N\xi)_{\mu\nu}&=&
2g_{\lambda(\nu}\nabla_{\mu)} \xi^\lambda\,,
\\[10pt]
%\eea
%\bea
(\bar Nh)^\alpha
%&=&
%-2\gamma^{\alpha\beta}E^{\rho\sigma\mu\nu}g_{\beta\rho}
%\nabla_{\sigma} h_{\mu\nu}
%\nonumber\\[5pt]
&=&-2\frac{\alpha}{k^2}
\left(g^{\alpha(\nu}\nabla^{\mu)}
-\varkappa g^{\mu\nu}\nabla^\alpha\right)h_{\mu\nu}
\,.
\eea
Therefore, the ghost operator $\tilde F=\bar NN:
C^{\infty}(TM)\to C^\infty(TM)$ 
is a second-order differential operator defined by 
\be
\tilde F=2\frac{\alpha}{k^2}F\,,
\ee
where
\be
F^\mu{}_\nu=-\delta^{\mu}{}_{\nu}\Delta
+(2\varkappa-1)\nabla^\mu\nabla_\nu
-R^{\mu}{}_{\nu}\;.
\ee

The second variation of the action defines a second-order
partial differential operator 
$P: C^{\infty}({\cal T}_{(2)})\to C^\infty({\cal T}_{(2)})$ by
\be
\frac{d^2}{d\varepsilon^2}S_{GR}(g+\varepsilon h)\Big|_{\varepsilon=0}
=(h,Ph)_{{\cal T}_{(2)}}
\,,
\ee
where
\bea
P_{\mu\nu}{}^{\alpha\beta}
&=&
-{1\over 2 k^2}\Biggl\{
-\left(\delta^{(\alpha}{}_{(\mu}\delta^{\beta)}{}_{\nu)}
+\frac{1-2\varkappa}{n\varkappa-1}
g_{\mu\nu}g^{\alpha\beta}\right)\Delta
%\nonumber\\[10pt]
%&&
+\frac{1-2\varkappa}{n\varkappa-1}
g_{\mu\nu}\nabla^{(\alpha}\nabla^{\beta)}
\nonumber\\[10pt]
&&
-g^{\alpha\beta}\nabla_{(\mu}\nabla_{\nu)}
+2\nabla_{(\mu}\delta^{(\alpha}{}_{\nu)}\nabla^{\beta)}
-2R^\alpha{}_{(\mu}{}^\beta{}_{\nu)}
-2\delta^{(\alpha}{}_{(\mu}R^{\beta)}{}_{\nu)}
+R_{\mu\nu}g^{\alpha\beta}
\nonumber\\[10pt]
&&
+\frac{4\varkappa-1}{n\varkappa-1}
g_{\mu\nu}R^{\alpha\beta}
+(R-2\Lambda)\delta^{\alpha}{}_{(\mu}\delta^\beta{}_{\nu)}
\nonumber\\[10pt]
&&
+\frac{1}{2(n\varkappa-1)}\Big[(1-4\varkappa)R
+2(2\varkappa-1)\Lambda\Big]
g_{\mu\nu}g^{\alpha\beta}
\Biggr\}\,.
\eea
The operator 
$N\bar N: C^{\infty}({\cal T}_{(2)})\to C^\infty({\cal T}_{(2)})$
is a second-order operator of the form
\be
(N\bar N)_{\mu\nu}{}^{\alpha\beta}
=-4\frac{\alpha}{k^2}\left\{
\nabla_{(\mu}\delta^{(\alpha}{}_{\nu)}\nabla^{\beta)}
-\varkappa g^{\alpha\beta}\nabla_{(\mu}\nabla_{\nu)}
\right\}\,.
\ee
The graviton operator $\tilde L=-P- N\bar N:
C^{\infty}({\cal T}_{(2)})\to C^\infty({\cal T}_{(2)})$ now reads
\be
\tilde L=\frac{1}{2k^2}L\,,
\ee
where
\bea
L_{\mu\nu}{}^{\alpha\beta}
&=&
-\left(\delta^{(\alpha}{}_{(\mu}\delta^{\beta)}{}_{\nu)}
+\frac{1-2\varkappa}{n\varkappa-1}
g_{\mu\nu}g^{\alpha\beta}\right)\Delta
%\nonumber\\[10pt]
%&&
%+8\alpha\left\{
%\nabla_{(\mu}\delta^{(\alpha}{}_{\nu)}\nabla^{\beta)}
%-\varkappa g^{\alpha\beta}\nabla_{(\mu}\nabla_{\nu)}
%\right\}
+\frac{1-2\varkappa}{n\varkappa-1}
g_{\mu\nu}\nabla^{(\alpha}\nabla^{\beta)}
\nonumber\\[10pt]
&&
-(1+8\alpha\varkappa)
g^{\alpha\beta}\nabla_{(\mu}\nabla_{\nu)}
%\nonumber\\[10pt]
%&&
+2(1+4\alpha)\nabla_{(\mu}\delta^{(\alpha}{}_{\nu)}\nabla^{\beta)}
-2R^\alpha{}_{(\mu}{}^\beta{}_{\nu)}
\nonumber\\[10pt]
&&
-2\delta^{(\alpha}{}_{(\mu}R^{\beta)}{}_{\nu)}
+R_{\mu\nu}g^{\alpha\beta}
%\nonumber\\[10pt]
%&&
+\frac{4\varkappa-1}{n\varkappa-1}
g_{\mu\nu}R^{\alpha\beta}
+(R-2\Lambda)\delta^{\alpha}{}_{(\mu}\delta^\beta{}_{\nu)}
\nonumber\\[10pt]
&&
+\frac{1}{2(n\varkappa-1)}\Big[(1-4\varkappa)R
+2(2\varkappa-1)\Lambda\Big]
g_{\mu\nu}g^{\alpha\beta}\,.
\eea

The most convenient choice of the parameters of the fiber metrics
(gauge parameters) is
\be
\varkappa=\frac{1}{2}\,, \qquad
\alpha=-\frac{1}{4}
\,.
\ee
In this gauge the non-diagonal derivatives in both the operators $F$ and 
$L$ vanish so that they both are of Laplace type
\bea
F^{\mu}{}_{\nu}&=&-\delta^{\mu}{}_{\nu}\Delta-R^{\mu}{}_{\nu}\;,
\nonumber\\[10pt]
%\ee
%\be
L&=&-\Delta+Q\,,
\eea
where
\bea
Q_{\mu\nu}{}^{\alpha\beta}
&=&
-2R_\mu{}^{(\alpha}{}_\nu{}^{\beta)}
-2\delta^{(\alpha}{}_{(\mu}R^{\beta)}{}_{\nu)}
+R_{\mu\nu}g^{\alpha\beta}
+\frac{2}{n-2}g_{\mu\nu}R^{\alpha\beta}
\nonumber\\
&&
-\frac{1}{(n-2)}g_{\mu\nu}g^{\alpha\beta}R
+(R-2\Lambda)\delta^\alpha{}_{(\mu}\delta^\beta{}_{\nu)}
\,.
\eea

In the Euclidean formulation the zeta-regularized
effective action has the form
\be
\Gamma^{GR}_{(1)}=-\frac{1}{2}\zeta'_{GR}(0)\,.
\ee
where 
\be
\zeta_{GR}(s)=\zeta_{L}(s)-2\zeta_F(s)\,,
\ee
$\zeta_{L}(s)$ and $\zeta_F(s)$ are the zeta
functions of the graviton operator $L$ and the ghost
operator $F$.
Next, by using the definition of the zeta function we obtain
\be
\zeta_{GR}(s)=\frac{\mu^{2s}}{\Gamma(s)}
\int\limits_0^\infty dt\; t^{s-1}e^{t\lambda}
\Theta_{GR}(t)\,,
\ee
where 
\be
\Theta_{GR}(t)=
\Theta_{L}(t)
-2\Theta_F(t)\,,
\ee
and $\Theta_{L}(t)$ and $\Theta_F(t)$ are the heat
traces of the operators $L$ and $F$.

By using the results for the heat traces described above we
obtain the total heat trace
\bea
\Theta_{GR}(t)
&=&(4\pi t)^{-n/2}
\int\limits_M d\vol\;
\exp\left\{\left({1\over 8} R + {1\over 6} R_H\right)t\right\}
%\nonumber
%\label{437axxz}
\\
&&
\times
\int\limits_{\RR^n_{\rm reg}} 
\frac{d\omega}{(4\pi t)^{p/2}}\;\beta^{1/2}
\exp\left\{-{1\over 4 t}\left<\omega,\beta\omega\right>\right\}
\Psi_{GR}(t,\omega)
\nonumber\\
&& 
\times
\left[\det{}_\mathcal{H}
\left({\sinh\left[\,F(\omega)/2\right]\over 
\,F(\omega)/2}\right)\right]^{1/2}
\left[\det{}_{TM}\left({\sinh\left[\,D(\omega)/2\right]\over 
\,D(\omega)/2}\right)\right]^{-1/2}\,,
%\nonumber
\eea
where
\be
\Psi_{GR}(t,\omega)
=\Psi_L(t,\omega)-2\Psi_F(t,\omega)\,.
\ee
Thus, all we need to compute is 
the functions $\Psi_L(t,\omega)$ 
and $\Psi_F(t,\omega)$ for the operators $L$ and $F$.

Notice that both operators $L$ and $F$ act on pure (untwisted) tensor bundles.
Therefore, there is no Yang-Mills group, that is, ${\cal F}_{ab}={\cal
E}_{ab}={\cal B}_{ab}=0$. 
The generators of the orthogonal algebra ${\cal SO}(n)$ in the
vector and the
symmetric $2$-tensor
representation are
\bea
\left(\Sigma_{(1)}{}_{ab}\right)^c{}_d
&=&
2\delta^{c}{}_{[a}g_{b]d}\,,
%\nonumber
\\[5pt]
%\ee
%\be
\left(\Sigma_{(2)}{}_{ab}\right){}_{cd}{}^{ef}
&=&
-4\delta^{(e}{}_{[a}g_{b](d}\delta^{f)}{}_{c)}
\,.
\eea
Therefore, the generators of the holonomy group are
\bea
{\cal R}_{(1)}{}_i
&=&D_{i}\,,
\\[5pt]
%\ee
%and
%\be
{\cal R}_{(2)}{}_i
&=&
-2 D_{i}\vee I_{(1)}\,,
\eea
which, in component language, reads
\bea
\left({\cal R}_{(1)}{}_i\right)^a{}_b
&=&
D^a{}_{ib}\,,
\\[5pt]
%\ee
%and
%\be
\left({\cal R}_{(2)}{}_i\right)_{cd}{}^{ab}
&=&
-2 D^{(a}{}_{i(d}\delta^{b)}{}_{c)}\,.
\eea
Now, it is easy to compute the Casimir operators
\bea
\left({\cal R}^2_{(1)}\right)^a{}_b
&=&-R^a{}_{b}\,,
\\[5pt]
%\ee
%and
%\be
\left({\cal R}^2_{(2)}\right)_{cd}{}^{ab}
&=&
2R^{(a}{}_{d}{}^{b)}{}_{c}
-2\delta^{(a}{}_{(c}R^{b)}{}_{d)}
\,.
\eea
The potentials for both operators are obviously 
read off from their definition
\bea
\left(Q_{F}\right)^a{}_b
&=&-R^a{}_b\,,
%\nonumber
\\[5pt]
%\ee
%\bea
\left(Q_{L}\right)_{cd}{}^{ab}
&=&
-2R^{(a}{}_c{}^{b)}{}_d
-2\delta^{(a}{}_{(c}R^{b)}{}_{d)}
+R_{cd}g^{ab}
+\frac{2}{n-2}g_{cd}R^{ab}
\nonumber\\
&&
-\frac{1}{(n-2)}g_{cd}g^{ab}R
+\delta^a{}_{(c}\delta^b{}_{d)}(R-2\Lambda)
\,.
\eea

By substituting these expressions in the general formula
(\ref{314xxz}) we obtain 
\bea
\Psi_{L}(t,\omega)
&=&\exp\left[-t(R-2\Lambda)\right]\tr_{{\cal T}_{(2)}}
\exp\left(t{\cal M}_{L}\right)
\exp\left[2D(\omega)\vee I_{(1)}\right]\,,
\nonumber\\[10pt]
\Psi_F(t,\omega)
&=&
\tr_{TM}
\exp\left(t{\cal M}_{F}\right)
\exp\left[\,D(\omega)\right]
\,,
%\label{psixxx}
\eea
where the endomorphisms ${\cal M}_L$ and ${\cal M}_F$
are defined by
\bea
\left({\cal M}_{F}\right)^a{}_b
&=&
2R^a{}_b\,,
\\[5pt]
%\ee
%\be
\left({\cal M}_{L}\right)_{cd}{}^{ab}
&=&
4\delta^{(a}{}_{(c}R^{b)}{}_{d)}
-R_{cd}g^{ab}
-\frac{2}{n-2}g_{cd}R^{ab}
%\nonumber\\
%&&
+\frac{1}{(n-2)}g_{cd}g^{ab}R
\,.
\eea

%===============================================
\section{Yang-Mills Theory in Curved Space}
\setcounter{equation}0

Let $G_{YM}$ be a compact simple Lie group.
Yang-Mills theory is a gauge theory with
the gauge group being the group of transformations of
sections of the principal bundle over the 
spacetime manifold $M$ with structure group $G_{YM}$
and the configuration space being the space
of all connections on this principal bundle valued in the Lie algebra
${\cal G}_{YM}$ of the group $G_{YM}$.
Let $Ad: {\cal G}_{YM}\to \End(W_{Ad})$ be the adjoint representation 
of the gauge algebra ${\cal G}_{YM}$ in the vector space $W_{Ad}$
and ${\cal W}_{Ad}$ be the associated vector bundle over $M$
with structure group $G_{YM}$ and the fiber $\End(W_{Ad})$ realizing
the adjoint representation of the gauge group.
The Yang-Mills gauge field can be para\-met\-rized by the 
local components of the connection ${\cal A}_\mu$
taking values in $\End(W_{Ad})$.
Then the Yang-Mills action has the form  
\cite{dewitt03}
\bea
S_{YM} &=&\frac{1}{8e^2}\int\limits_M dx\;g^{1/2}
\tr_{W_{Ad}} g^{\mu\alpha}g^{\nu\beta}
{\cal F}_{\mu\nu}{\cal F}_{\alpha\beta}\,.
\eea

The (ghost) operator $K: C^\infty({\cal W}_{Ad})\to C^\infty({\cal W}_{Ad})$
and the (gluon) operator $H: C^\infty({\cal W}_{Ad}\otimes TM)\to
C^\infty({\cal W}_{Ad}\otimes TM)$
are second-order partial differential operators
acting on scalar and 
vector fields valued in $\End(W_{Ad})$.
In the minimal gauge these operators are
\cite{avramidi95b,avramidi99}
\bea
H{}^{\mu}{}_{\nu}
&=&-\delta^\mu{}_\nu\Delta+R^{\mu}{}_{\nu}
-2{\cal F}^{\mu}{}_{\nu}\,,
\\[10pt]
K&=&-\Delta\,.
\eea

Thus, the zeta-regularized one-loop effective action
of quantum Yang-Mills theory in the Euclidean formulation
is given by
\be
\Gamma^{YM}_{(1)}=-\frac{1}{2}\zeta'_{YM}(0)\,.
\ee
where 
\be
\zeta_{YM}(s)=\zeta_{H}(s)-2\zeta_{K}(s)\,,
\ee
$\zeta_{H}(s)$ and $\zeta_{K}(s)$ are the zeta
functions of the gluon operator $H$ and the ghost
operator $K$.
Next, by using the definition of the zeta function we obtain
\be
\zeta_{YM}(s)=\frac{\mu^{2s}}{\Gamma(s)}
\int\limits_0^\infty dt\; t^{s-1}e^{t\lambda}
\Theta_{YM}(t)\,,
\ee
where 
\be
\Theta_{YM}(t)=
\Theta_{H}(t)
-2\Theta_{K}(t)\,,
\ee
$\Theta_{H}(t)$ and $\Theta_{K}(t)$ are the heat
traces of the operators $H$ and $K$.

Since both operators $H$ and $K$ are of Laplace type we can use the results
for the heat trace described above. We obtain
the total heat trace
\bea
\Theta_{YM}(t)
&=&(4\pi t)^{-n/2}
\int\limits_M dx\,g^{1/2}\;
\exp\left\{\left({1\over 8} R + {1\over 6} R_H\right)t\right\}
%\nonumber
%\label{437axxz}
\\
&&
\times
\int\limits_{\RR^n_{\rm reg}} 
\frac{d\omega}{(4\pi t)^{p/2}}\;\beta^{1/2}
\exp\left\{-{1\over 4 t}\left<\omega,\beta\omega\right>\right\}
\Psi_{YM}(t,\omega)
\nonumber\\
&& 
\times
\left[\det{}_\mathcal{H}
\left({\sinh\left[\,F(\omega)/2\right]\over 
\,F(\omega)/2}\right)\right]^{1/2}
\left[\det{}_{TM}\left({\sinh\left[\,D(\omega)/2\right]\over 
\,D(\omega)/2}\right)\right]^{-1/2}
%\nonumber
\eea
where
\bea
\Psi_{YM}(t,\omega)
&=&\Psi_H(t,\omega)-2\Psi_K(t,\omega)\,.
%\label{psixxx}
\eea
Thus all we have to do now is to compute the functions
$\Psi_H(t,\omega)$ and $\Psi_K(t,\omega)$ for the operators $H$ and $K$.

We assume that the gauge algebra is big enough to include the holonomy algebra
as a subalgebra (as discussed above). Further, we assume that ${\cal B}_{ab}$
takes values in the (Abelian) Cartan subalgebra of the gauge algebra. The other
part ${\cal E}_{ab}$ of the Yang-Mills curvature is described by a
representation $Y_{Ad}: {\cal SO}(n)\to \End(W_{Ad})$ of the orthogonal algebra
${\cal SO}(n)$ in $W_{Ad}$
with generators $Y^{Ad}_{ab}$ so that the total Yang-Mills
curvature is given by (\ref{2153mm})
\bea
Ad(\mathcal{F}_{ab})&=&
\frac{1}{2}R^{cd}{}_{ab}Y^{Ad}_{cd}
+Ad(\mathcal{B}_{ab})
\,,
\label{2153mm}
\eea

For the ghost 
operator $K$ we have $Q_K=0$ and $\Sigma^K_{ab}=0$. Therefore,
\be
{\cal R}^K_i=-\frac{1}{2}D^a{}_{ib}Y_{Ad}{}^b{}_a\,,
\ee
\be
{\cal R}^2_K=\frac{1}{4}R^{abcd}Y^{Ad}_{ab}Y^{Ad}_{cd}\,.
\ee
Thus, we obtain
\bea
\Psi_K(t,\omega)
&=&
\tr_{W_{Ad}}
\left[\det{}_{TM}\left(\frac{
\sinh(tAd(\mathcal{B}))}{tAd(\mathcal{B})}\right)\right]^{-1/2}
\nonumber\\[10pt]
&&\times
\exp\left(-{\cal R}^2_Kt\right)
%\nonumber\\[10pt]
%&&\times
\exp\left[{\cal R}_K(\omega)\right]\,,
\eea
where ${\cal R}_K(\omega)={\cal R}^K_i\omega^i$.

For the gluon 
operator $H$ we have
\bea
\left(Q_H\right)_{}^a{}_{b}
&=&
R^a{}_b
-2Ad({\cal F}^a{}_b)
\nonumber\\[5pt]
&=&
R^a{}_b
-R^a{}_{bcd}Y_{Ad}^{cd}-2Ad(\mathcal{B}^a{}_b)
\,,
\eea
\bea
\left(\Sigma^H_{}{}_{ab}\right)^c{}_d
&=&
2\delta^{c}{}_{[a}g_{b]d}\,.
\eea
Therefore,
\bea
\left({\cal R}^H_{}{}_i\right)^a{}_b
&=&D^a{}_{ib}+\delta^a{}_b{\cal R}^K_i
\,,
\eea
and,
\bea
\left({\cal R}^2_{H}\right)^a{}_b
&=&
-R^a{}_b+R^a{}_{bcd}Y_{Ad}^{cd}
+\delta^a{}_b{\cal R}^2_K
\,,
\eea
Thus
\bea
\Psi_H(t,\omega)&=&
\tr_{W_{Ad}}
\left[\det{}_{TM}\left(\frac{
\sinh(tAd(\mathcal{B}))}{tAd(\mathcal{B})}\right)\right]^{-1/2}
\exp\left(-{\cal R}^2_Kt\right)
\nonumber\\[10pt]
&&
\times
\exp\left[{\cal R}_K(\omega)\right]
\tr_{TM}
\exp\left[2Ad({\cal B})t\right]
\exp\left[D(\omega)\right]\,.
\eea

Finally, we obtain the total function $\Psi(t,\omega)$:
\bea
\Psi_{YM}(t,\omega)&=&
\tr_{W_{Ad}}
\left[\det{}_{TM}\left(\frac{
\sinh(tAd(\mathcal{B}))}{tAd(\mathcal{B})}\right)\right]^{-1/2}
\exp\left(-{\cal R}^2_Kt\right)
\exp\left[{\cal R}_K(\omega)\right]
\nonumber\\[10pt]
&&
\times
\tr_{TM}
\Bigl\{
\exp\left[2Ad({\cal B})t\right]
\exp\left[D(\omega)\right]
-2\Bigr\}\,.
\nonumber\\
\eea

%====================================================
\section{Matter Fields}

%=====================================================

Now, we assume that $M$ is a spin manifold. Let $\Lambda_{\rm spin}$ be the
spinor vector space and $\End(\Lambda_{\rm spin})$ be the space of
endomorphisms of $\Lambda_{\rm spin}$. Let $\mathcal{S}$ be the spinor bundle
with fiber $\Lambda_{\rm spin}$ realizing the spinor representation of the spin
group ${\rm Spin}(n)$. It defines the spinor representation $\gamma: {\cal
SO}(n)\to \End(\Lambda_{\rm spin})$ of the orthogonal algebra ${\cal SO}(n)$ in
$\Lambda_{\rm spin}$.
The spin connection induces a connection on the bundle $\mathcal{S}$ defining
the covariant derivative of spin-tensor fields.
Let $G_{YM}$ be a  compact simple Lie group and ${\cal G}_{YM}$ be its Lie
algebra. It naturally defines the principal fiber bundle over the manifold $M$
with the structure group $G_{YM}$. Let $W_{\rm spin}$ be a vector space and
$\End(W_{\rm spin})$ be the space of its endomorphisms. We consider a
representation $X_{\rm spin}: {\cal G}_{YM}\to \End(W_{\rm spin})$ of the Lie
algebra ${\cal G}_{YM}$ in $W_{\rm spin}$ and the associated vector bundle
${\cal W}_{\rm spin}$ through this representation with the same structure group
$G_{YM}$ whose typical fiber is $W_{\rm spin}$. Then we define the twisted
spinor bundle $\mathcal{V}_{\rm spin}$ via the twisted product of the bundles
$\mathcal{W}_{\rm spin}$ and $\mathcal{S}$ with the fiber $V_{\rm
spin}=\Lambda_{\rm spin}\otimes W_{\rm spin}$. The spin connection on the
spinor bundle and the Yang-Mills connection on the bundle ${\cal W}_{\rm spin}$
define the twisted spin connection on the bundle ${\cal V}_{\rm spin}$.

Let ${\cal W}_{0}$ be another associated vector bundle over $M$ with the
structure group $G_{YM}$ and typical fiber $W_{0}$ realizing a representation
$X_{0}: {\cal G}_{YM}\to \End(W_{0})$ of the Lie algebra ${\cal G}_{YM}$ in
$W_{0}$. 

The sections of the bundles ${\cal W}_0$ and ${\cal V}_{\rm spin}$
are multiplets of scalar, $\varphi$, and spinor, $\psi$, 
fields that we call
matter fields.
The action of matter fields reads
\bea
S_{\rm matter}&=&\int\limits_M dx\,g^{1/2}\Biggl\{
\left<\psi,\left[\gamma^\mu\nabla_\mu
+M(\varphi)\right]\psi\right>_{V_{\rm spin}} 
\nonumber\\[10pt]
&&
-{1\over 2}g^{\mu\nu}\left<\nabla_\mu\varphi,\nabla_\nu\varphi\right>_{W_0}
-V(\varphi)\Biggr\},
%\label{(1)}
\eea
where $\left<\;,\;\right>_{V_{\rm spin}}$
and $\left<\;,\;\right>_{W_0}$ are the (Hermitian) 
inner products in the vector spaces
$V_{\rm spin}$ and $W_0$, $M(\varphi)\in \End(V_{\rm spin})$ is an endomorphism
of $V_{\rm spin}$ and $V(\varphi)$ is a scalar function of $\varphi$.

The contribution of the matter fields to the one-loop effective action has the
form \cite{dewitt03}
\begin{equation}
\Gamma^{\rm matter}_{(1)}=-\log\Det\,D+{1\over 2}\log\Det\,L_0\,,
\label{(9)}
\end{equation}
where $D$ is the Dirac type operator and $L_0$ is a Laplace type operator
defined by
\be
D=\gamma^\mu\nabla_\mu+M(\phi)\,,
\ee
\begin{equation}
L_0 = -\Delta +Q_0(\phi),
\label{(10)}
\end{equation}
$\phi$ is a background scalar field and $Q_0(\phi)$ is 
the mass matrix of the scalar fields.
Here the background scalar fields realize the minimum of the potential 
$V(\varphi)$, and the matrix $Q_0$ is defined by
\be
V(\varphi)=V(\phi)
+\frac{1}{2}\left<(\varphi-\phi),Q_0(\phi)(\varphi-\phi)\right>_{W_0}
+O((\varphi-\phi)^3)\,.
\ee
As we mentioned above it is assumed that the endomorphism $Q_0$ is covariantly
constant.

We also assume that the mass matrix $M$ does not
contain the Dirac matrices or contains only even number of them, so
that 
\be
[M,\gamma_\mu]=0
\ee
Then one can show that the spinor contribution
can be expressed in terms of the squared Dirac operator
\begin{equation}
\log\Det\,D={1\over 2}\log\Det\,L_{\rm spin}\,,
\label{(14)}
\end{equation}
where
\begin{eqnarray}
L_{\rm spin}&&=-\Delta+\frac{1}{4}R
-{1\over 2}\gamma^{ab}X({\cal F}_{ab}) 
+M^2\,.
\label{(15)}
\end{eqnarray}
where 
$
\gamma_{ab}=\gamma_{[a}\gamma_{b]}\,.
$

Thus, the zeta-regularized contribution of the matter fields to the
one-loop effective action
in the Euclidean formulation
is given by
\be
\Gamma^{\rm matter}_{(1)}=-\frac{1}{2}\zeta'_{\rm matter}(0)\,.
\ee
where 
\be
\zeta_{\rm matter}(s)=\zeta_{0}(s)-\zeta_{\rm spin}(s)\,,
\ee
$\zeta_{0}(s)$ and $\zeta_{\rm spin}(s)$ are the zeta
functions of the operators $L_0$ and 
$L_{\rm spin}$.
Next, by using the definition of the zeta function we obtain
\be
\zeta_{\rm matter}(s)=\frac{\mu^{2s}}{\Gamma(s)}
\int\limits_0^\infty dt\; t^{s-1}e^{t\lambda}
\Theta_{\rm matter}(t)\,,
\ee
where 
\be
\Theta_{\rm matter}(t)=
\Theta_{0}(t)
-\Theta_{\rm spin}(t)\,,
\ee
$\Theta_{0}(t)$ and $\Theta_{\rm spin}(t)$ are the heat
traces of the operators $L_0$ and $L_{\rm spin}$.

Since both operators $L_0$ and $L_{\rm spin}$ are of Laplace type we can use
the results
for the heat trace described above. We obtain
the total heat trace
\bea
\Theta_{\rm matter}(t)
&=&(4\pi t)^{-n/2}
\int\limits_M dx\,g^{1/2}\;
\exp\left\{\left({1\over 8} R + {1\over 6} R_H\right)t\right\}
\nonumber
%\label{437axxz}
\\
&&
\times
\int\limits_{\RR^n_{\rm reg}} 
\frac{d\omega}{(4\pi t)^{p/2}}\;\beta^{1/2}
\exp\left\{-{1\over 4 t}\left<\omega,\beta\omega\right>\right\}
\Psi_{\rm matter}(t,\omega)
\nonumber\\
&& 
\times
\left[\det{}_\mathcal{H}
\left({\sinh\left[\,F(\omega)/2\right]\over 
\,F(\omega)/2}\right)\right]^{1/2}
\left[\det{}_{TM}\left({\sinh\left[\,D(\omega)/2\right]\over 
\,D(\omega)/2}\right)\right]^{-1/2}
\label{734xxc}
\eea
where
\bea
\Psi_{\rm matter}(t,\omega)
&=&\Psi_{0}(t,\omega)-\Psi_{\rm spin}(t,\omega)\,.
%\label{psixxx}
\eea
Thus all we have to do now is to compute the functions
$\Psi_{0}(t,\omega)$ and $\Psi_{\rm spin}(t,\omega)$ for the operators 
$L_0$ and
$L_{\rm spin}$.

We assume that the gauge algebra is big enough to include the holonomy algebra
as a subalgebra (as discussed above). Further, we assume that ${\cal B}_{ab}$
takes values in the (Abelian) Cartan subalgebra of the gauge algebra and 
${\cal E}_{ab}$ takes values in the corresponding repesentation of the
holonomy algebra. More precisely, we define two representations of the
orthogonal algebra $Y_{0}: {\cal SO}(n)\to \End(W_{0})$
and $Y_{\rm spin}: {\cal SO}(n)\to \End(W_{\rm spin})$
with generators $Y^{0}_{ab}$ and $Y^{\rm spin}_{ab}$ so that the total
Yang-Mills curvature in the representations $X_0$ and $X_{\rm spin}$ is given
by (\ref{2153mm})
\bea
X_0(\mathcal{F}_{ab})&=&
\frac{1}{2}R^{cd}{}_{ab}Y^{0}_{cd}
+X_0(\mathcal{B}_{ab})
\\[10pt]
X_{\rm spin}(\mathcal{F}_{ab})&=&
\frac{1}{2}R^{cd}{}_{ab}Y^{\rm spin}_{cd}
+X_{\rm spin}(\mathcal{B}_{ab})
\,.
%\label{2153mm}
\eea

Now, for the  scalar
operator $L_0$ we have $\Sigma^0_{ab}=0$, and, therefore,
\be
{\cal R}^0_i=-\frac{1}{2}D^a{}_{ib}Y_0{}^b{}_a\,,
\ee
\be
{\cal R}^2_0=\frac{1}{4}R^{abcd}Y^0_{ab}Y^0_{cd}\,.
\ee
Thus, we obtain
\bea
\Psi_0(t,\omega)
&=&
\tr_{W_0}
\left[\det{}_{TM}\left(\frac{
\sinh(tX_0(\mathcal{B}))}{tX_0(\mathcal{B})}\right)\right]^{-1/2}
\exp\left[-\left(\frac{1}{4}R^{abcd}Y^0_{ab}Y^0_{cd}+Q_0\right)t\right]
\nonumber\\[10pt]
&&\times
\exp\left[-\frac{1}{2}D^a{}_{ib}Y_0{}^b{}_a\omega^i\right]\,.
\eea

For the spinor
operator $L_{\rm spin}$ we have
\bea
Q_{\rm spin}
&=&\frac{1}{4}R-\frac{1}{2}\gamma^{ab}X_{\rm spin}({\cal F}_{ab})+M^2
\nonumber\\[5pt]
&=&
\frac{1}{4}R
-\frac{1}{4}R^{abcd}\gamma_{ab}Y^{\rm spin}_{cd}
-\frac{1}{2}\gamma^{ab}X_{\rm spin}({\cal B}_{ab})
+M^2
\,,
\eea
The generators of the orthogonal algebra in the spinor representation are
\bea
\Sigma^{\rm spin}_{ab}
&=&\frac{1}{2}\gamma_{ab}\,.
\eea
Therefore,
\bea
{\cal R}^{\rm spin}_i
&=&-\frac{1}{2}D^a{}_{ib}\left(\gamma^b{}_a
+Y_{\rm spin}{}^b{}_{a}\right)\,,
\eea
and
\bea
{\cal R}^2_{\rm spin}
&=&
-\frac{1}{8}R
+\frac{1}{4}R^{abcd}\gamma_{ab}Y^{\rm spin}_{cd}
+\frac{1}{4}R^{abcd}Y^{\rm spin}_{ab}Y^{\rm spin}_{cd}
\,,
\eea
Thus the endomorphism ${\cal R}_{\rm spin}^2+Q_{\rm spin}$ has the form
\be
{\cal R}_{\rm spin}^2+Q_{\rm spin}=
\frac{1}{8}R
+\frac{1}{4}R^{abcd}Y^{\rm spin}_{ab}Y^{\rm spin}_{cd}
-\frac{1}{2}\gamma^{ab}X_{\rm spin}({\cal B}_{ab})
+M^2
\,.
\ee
Finally, we obtain
\bea
\Psi_{\rm spin}(t,\omega)
&=&
\exp\left(-\frac{1}{8}R t\right)
\tr_{W_{\rm spin}}
\left[\det{}_{TM}\left(\frac{\sinh(tX_{\rm spin}(\mathcal{B}))}
{tX_{\rm spin}(\mathcal{B})}\right)\right]^{-1/2}
\nonumber\\[10pt]
&&
\times
\exp\left[-\left(\frac{1}{4}R^{abcd}Y^{\rm spin}_{ab}Y^{\rm spin}_{cd}
+M^2\right)
t\right]
%\nonumber\\[10pt]
%&&
\exp\left[-\frac{1}{2}D^a{}_{ib}\omega^iY_{\rm spin}{}^b{}_a\right]
\nonumber\\[10pt]
&&
\times
\tr_{\Lambda_{\rm spin}}
\exp\left[-\frac{1}{2}\left(X_{\rm spin}({\cal B}^a{}_b)t
+D^a{}_{ib}\omega^i\right)\gamma^b{}_a\right]\,,
\eea
where $\tr_{\Lambda_{\rm spin}}$
indicates the trace over the spinor indices.
It is interesting to note that the scalar curvature term $\exp(-\frac{1}{8}R)$
in the function $\Psi_{\rm spin}(t,\omega)$ precisely cancels the prefactor 
$\exp(\frac{1}{8}R)$
in the heat trace (\ref{734xxc}).

%=====================================================
\section{Conclusion}

In the present paper we used the results for the heat kernel on homogeneous
bundles over symmetric spaces obtained in our recent paper \cite{avramidi08b}
by using sophisticated algebraic methods to evaluate the low-energy effective
action in quantum gravity and gauge (Yang-Mills) theory There always exists a
minimal gauge such that both the gauge field operator and the ghost operator
are of laplace type, and, therefore, the evaluation of the zeta-regularized
effective action reduces to the calculation of the corresponding heat traces.
Of course, one could try to go further and compute the functions
$\Psi(t;\omega)$ for the relevant operators by finding the eigenvalues of the
corresponding endomorphisms  etc.  However, we will not do this in this
paper and leave the result in the general form it was presented above.

We would like to stress two more points here. First of all, quantum general
relativity is a non-renormalizable theory. Therefeore, even if one gets a final
result via the zeta-regularization one should not take it too seriously.
Secondly, our results for the heat kernel and, hence, for the effective action
are essentially non-perturbative. They contain an infinite series of Feynmann
diagrams with low momenta and cannot be obtained in any perturbation theory.
One could try now to use this result for the analysis of the ground state in
quantum gravity. But this is a rather ambitious program for the future.

%=======================================
%=======================================
%=======================================

\end{document}